\documentclass[baaa]{baaa}

\usepackage[pdftex]{hyperref}
\usepackage{subfigure}
\usepackage{natbib}
\usepackage{helvet,soul}
\usepackage[font=small]{caption}

\begin{document}


\journalvol{61A}
\journalyear{2019}
\journaleditors{R. Gamen, N. Padilla, C. Parisi, F. Iglesias \& M. Sgr\'o}


\contriblanguage{1}


\contribtype{2}

\thematicarea{6}

\title{Studying star-forming processes towards G29.862--0.044}


\titlerunning{Studying star-forming processes}

\author{M. B. Areal\inst{1}, S. Paron\inst{1}, M. E. Ortega\inst{1} \& C. Fariña\inst{2}}

\authorrunning{Areal et al.}


\contact{mbareal@iafe.uba.ar}

\institute{Instituto de Astronomía y Física del Espacio, CONICET-UBA, Argentina \and Isaac Newton Group of Telescopes, La Palma, España and Instituto de Astrofísica de Canarias, La Laguna, España }


\resumen{Presentamos un estudio multiespectral hacia el objeto estelar joven (YSO, por su sigla en inglés) G29.862--0.044 (en adelante G29), localizado en la región de formación estelar G29.96--0.02, y ubicado a una distancia de aproximadamente 6.5 kpc.
Para estudiar el medio interestelar circundante a G29 se utilizaron datos de líneas moleculares (resolución angular de 15\arcsec) obtenidos de la base de datos del Telescopio James Clerk Maxwell.
Se caracterizó la nube molecular en la que se encuentra embebido el YSO y se determinaron parámetros físicos del {\it outflow} molecular relacionado. Adicionalmente se analizaron datos del infrarrojo cercano (resolución espacial del orden de 0.5\arcsec) obtenidos con el instrumento NIRI en Gemini Norte con el fin de observar más detalladamente el ambiente circumestelar de G29. La emisión en la banda Ks hacia G29 presenta una morfología en forma de cono que apunta hacia el outflow molecular rojo.
El objetivo de este trabajo es obtener un panorama lo más completo posible del YSO, de los procesos que generan la formación estelar, y del medio interestelar circundante. Estudios como este son de relevancia porque contribuyen a una comprensión integral de la formación de estrellas.}

\abstract{We present a multiwavelength study towards the young stellar object (YSO) G29.862--0.044 (hereafter G29), which is embedded in the massive star-forming region G29.96--0.02, located at a distance of about 6.5~kpc.
The surrounding interstellar medium of G29 is studied using molecular lines data (angular resolution about 15~\arcsec) obtained from the databases of the James Clerk Maxwell Telescope.
The physical conditions of G29 molecular outflows and the clump where the YSO is embedded are characterized. Near-IR data is also analyzed (spatial resolution of about 0.5~\arcsec) obtained with NIRI at Gemini North to have a detailed view of the circumstellar ambient of G29. The Ks emission towards G29 exhibits a cone-like feature pointing to the red molecular outflow.
The aim of this work is to obtain a complete picture of this YSO, the related star-forming processes, and the interstellar medium around it. Studies like this are important because they contribute to a comprehensive understanding of star formation.}



\maketitle

\section{Introduction}

\begin{figure}[!t]
  \centering
  \includegraphics[width=0.48\textwidth]{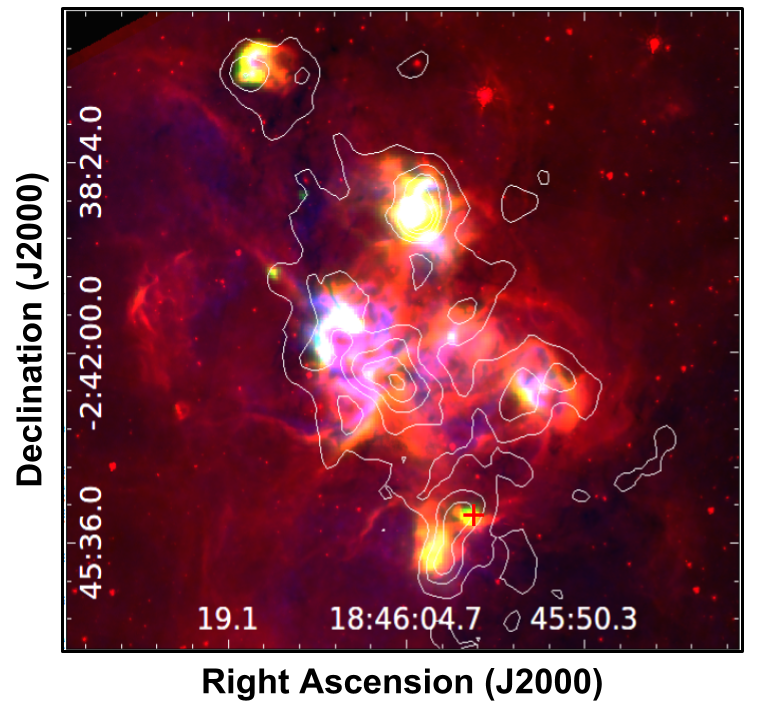}
  \caption{Three-color image towards G29.96--0.02 displaying the Spitzer-IRAC 8~$\mu m$ emission in red, the Herschel-PACS 70~$\mu m$ emission in green, and the radio continuum emission at 20~cm, as extracted from the MAGPIS, in blue. Contours of the C$^{18}$O J=3–2 line integrated between 96 and 106~kms$^{-1}$ with levels of 5, 11, 16, 22 and 27~K km s$^{-1}$ are included. The YSO position is indicated with the red cross.}
  \label{rgb}
\end{figure}

Young stellar objects (YSOs) are usually found embedded in dense molecular clumps, environments with plenty of molecular gas and interstellar dust. The region G29.9--0.02 (see Fig.~\ref{rgb}). located at the distance of 6.2~kpc, is very active in massive star formation. Figure~\ref{rgb} is a three-colour image displaying the Spitzer-IRAC 8~$\mu$m emission in red, the Herschel-PACS 70~$\mu$m emission in green, and in blue, the radio continuum emission at 20 cm as extracted from the New GPS of the Multi-Array Galactic Plane Imaging Survey (MAGPIS). The emissions were selected to remark the borders of the photodissociation regions (PDRs) (displayed in the 8~$\mu$m emission), the distribution of the ionized gas (shown with the 20~cm emission) and the warm dust (70~$\mu$m emission). In a previous work we studied the $^{13}$CO/C$^{18}$O abundance ratio through the whole region in relation with these sources and with star-forming processes \citep*{2018A&A...617A..14P}. In the present work we focus on the study of the southern massive YSO cataloged as the Red MSX source G029.862--0.044 (hereafter G29).

\section{Data}

In order to perform a multispectral study of G29, we used public data ($^{12}$CO, $^{13}$CO and C$^{18}$O J=3--2), and dedicated observations (near-IR data).

The data of the CO isotopes were obtained from public databases which were generated through observations from the 15~m James Clerck Maxwell Telescope. The $^{12}$CO J=3--2 data belong to the COHRS survey \citep{2013ApJS..209....8D}, while the data of the other CO isotope belong to the CHIMPS survey \citep{2016MNRAS.456.2885R}.

The near-IR observations were carried out with the Near InfraRed Imager and Spectrometer (NIRI; \citealt{hodapp03}) at Gemini-North 8.2-m telescope, on July 2017 in queue mode (Band-1 Program GN-2017B-Q25).
NIRI was used with the f/6 camera that provides a plate scale of 0$.\!\!\arcsec$117 pix$^{-1}$ in a field of view of 120$\arcsec\times$ 120$\arcsec$. The seeing of the Ks image presented here is about 0$.\!\!\arcsec$4.

\section{Results}
\subsection{Molecular cloud}

\begin{figure}[!t]
  \centering
  \includegraphics[width=0.45\textwidth]{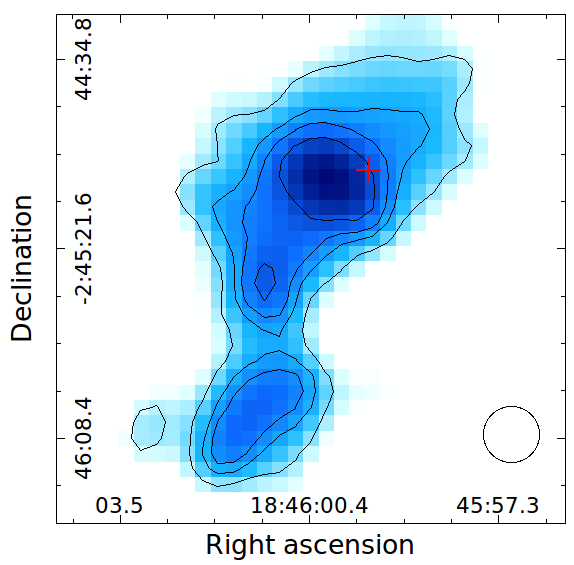}
  \caption{C$^{18}$O J=3--2 emission integrated between 96 and 106 km s$^{-1}$. The contour levels are 3, 5, 9 and 13~K km s$^{-1}$. The YSO position is indicated with the red cross and the beam is represented by the circle.}
  \label{mapaC18O}
\end{figure}

To characterize the molecular cloud in which the YSO is embedded we used the C$^{18}$O J=3--2 emission because it is an optically thin tracer of the molecular gas. The C$^{18}$O emission integrated between 96 and 106~km s$^{-1}$ is presented in Fig.~\ref{mapaC18O} showing the morphology of the molecular cloud associated with G29.

Assuming local thermodynamic equilibrium and using its typical formulae, the mass of the cloud was obtained. The H$_{2}$ column density, $N$(H$_{2}$), was calculated from the C$^{18}$O column density and using the conversion factor X$^{18}\sim 1.7 \times 10^{-7}$ \citep{1999RPPh...62..143W}. The estimated  mass is $M\sim 1\times 10^{4}$~M$_{\odot}$.

\subsection{Molecular outflow}

The YSOs at the earliest stages of formation are characterized by the presence of molecular outflows. We used the $^{12}$CO J=3--2 data to find the molecular outflows related 
to G29. We identified the presence of a red molecular outflow, but the blue molecular outflow can not be distinguished. This can be due to the presence of a dense concentration 
of cold dust southwards the YSO, AGAL G029.852--00.059 from \citet{urqu14}, that could hide the blue molecular outflow. Also the molecular gas related to the cold dust feature may confuse the observation of the gas associated with the blue outflow. This result is in agreement with that of \citet{2016AJ....152...92L}, who studied the outflowing activity of G29 using the $^{12}$CO J=1--0 line.

\begin{figure}[!t]
  \centering
  \includegraphics[width=0.48\textwidth]{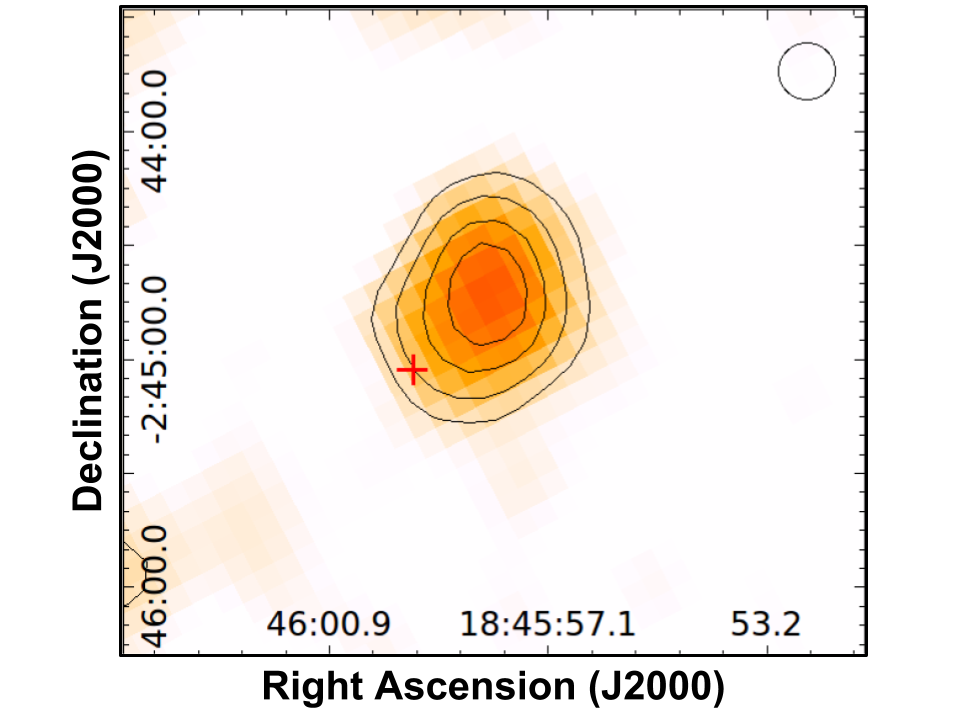}
  \caption{$^{12}$CO J=3--2 emission integrated between 108 and 113 km s$^{-1}$. The contour levels are 7, 11, 14 and 18~K km s$^{-1}$. The YSO position is indicated with the red cross and the beam is represented by the circle.}
  \label{mapa12CO}
\end{figure}

The $^{12}$CO emission integrated between 108 and 113~km s$^{-1}$ is presented in Fig.~\ref{mapa12CO}, which shows the morphology of the molecular outflow. We estimated the outflow physical parameters from the $^{12}$CO emission, obtaining a mass of $M_{\rm red} \sim $82$M_{\odot}$ and an energy $E_{\rm Kred}\sim2\times10^{46}$~erg. These are typical values of massive molecular outflows according to \citet{2015MNRAS.453..645M}.

\subsection{Cicumstellar ambient}

\begin{figure}[!t]
  \centering
  \includegraphics[width=0.48\textwidth]{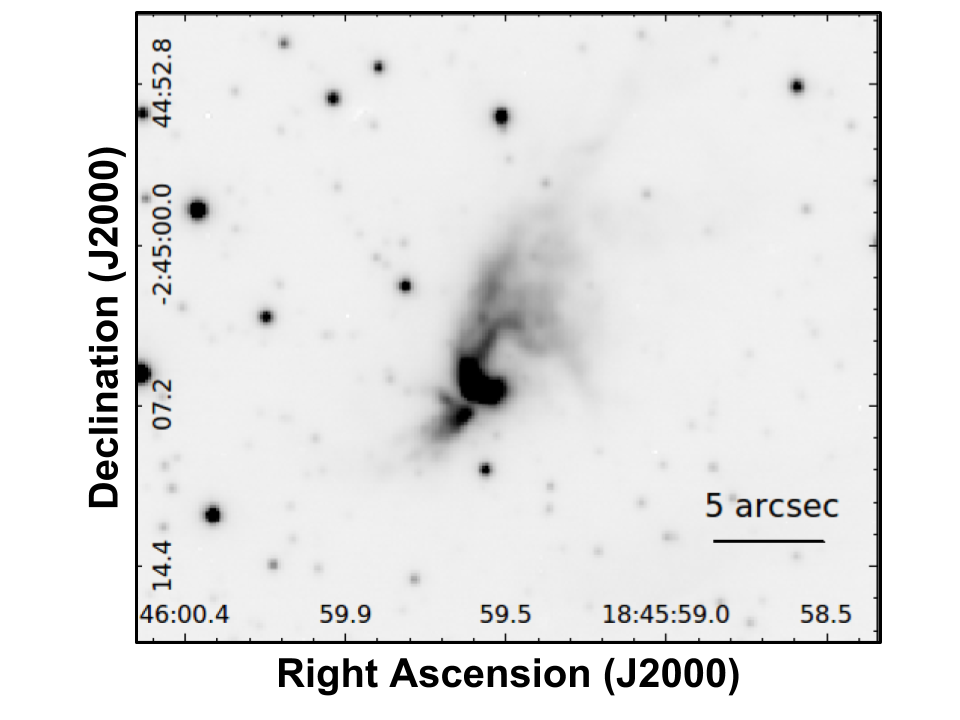}
  \caption{Ks emission obtained with NIRI-Gemini.}
  \label{bandaK}
\end{figure}

With the aim of studying in more detail the YSO and its surroundings we performed near-infrared observations using NIRI at Gemini North. The J, H and Ks broad bands, and some narrow bands such as lines of the molecular hydrogen, between others, were observed. In this work we present only the Ks emission (see Fig.~\ref{bandaK}).

The Ks emission towards G29,  exhibits a cone-like feature pointing to the northwest in direction to the red molecular outflow. 
It can be seen that there are two arc-like features inside the cone-like structure, which is very similar to what was found towards massive young stellar 
objects characterized by the presence of
precessing jets \citep{2006A&A...447..655W, 2013A&A...559L...2P, 2016A&A...593A.132P}. Also in Fig.~\ref{bandaK} it is observed a smaller structure towards the southwest,
which may be generated by a jet extending to the opposite direction.

As proposed in previous works (see for example \citealt{2006A&A...447..655W}) the near-IR emission at the Ks band towards this kind of objects likely arises from a cavity cleared in the circumstellar material by the action of a jet, and may be due to a combination of different emitting processes: radiation from the central protostar that is reflected at the inner walls of the cavity, the emission from warm dust, and the emission of some lines that these bands contain.
Further investigation is needed in order to determine such processes.
We are currently analyzing in detail the near-IR data set (the JHKs broad-bands and specially the emission lines observed with the narrow-bands), which will provide valuable information about the origin and physical processes taking place in this intriguing structure.

\bigskip

\noindent {\it Acknowledgements.}  This work was partially supported by grants awarded by ANPCYT and UBA (UBACyT) from Argentina. M.B.A. specially thanks the financial support received to assist to the Second Binational Meeting AAA-SOCHIAS.


\bibliographystyle{baaa}
\small
\bibliography{biblio}
 
\end{document}